\documentclass{INTERSPEECH2023}
\usepackage{multirow}
\usepackage[normalem]{ulem}
\useunder{\uline}{\ul}{}
\usepackage[caption=false]{subfig}
\usepackage{graphicx}
\usepackage{lipsum}  

\interspeechcameraready

\title{MCR-Data2vec 2.0: Improving Self-supervised Speech Pre-training via Model-level Consistency Regularization}
\name{Ji Won Yoon$^1$, Seok Min Kim$^1$, Nam Soo Kim$^1$}
\address{
  $^1$Department of ECE and INMC, Seoul National University, Seoul,
Republic of Korea}
\email{\{jwyoon, smkim\}@hi.snu.ac.kr, nkim@snu.ac.kr}

\begin{document}

\maketitle
 
\begin{abstract}
Self-supervised learning (SSL) has shown significant progress in speech processing tasks. However, despite the intrinsic randomness in the Transformer structure, such as dropout variants and layer-drop, improving the model-level consistency remains under-explored in the speech SSL literature. To address this, we propose a new pre-training method that uses consistency regularization to improve Data2vec 2.0, the recent state-of-the-art (SOTA) SSL model. Specifically, the proposed method involves sampling two different student sub-models within the Data2vec 2.0 framework, enabling two output variants derived from a single input without additional parameters. Subsequently, we regularize the outputs from the student sub-models to be consistent and require them to predict the representation of the teacher model. Our experimental results demonstrate that the proposed approach improves the SSL model's robustness and generalization ability, resulting in SOTA results on the SUPERB benchmark.
\end{abstract}

\section{Introduction}

Self-supervised learning (SSL) has shown great promise in speech processing \cite{hubert,w2v,wavlm,data2vec}.
It uses large amounts of unlabeled speech audio data to learn general speech representations, which can benefit various downstream tasks by fine-tuning. 

Recently, research in speech SSL algorithms has focused on data augmentation for a noisy scenario to improve the pre-training of SSL models \cite{wavlm, ccc-wav2vec, aug-cpc, w2v-aug, mvc, robust_data2vec}.
WavLM \cite{wavlm} employs a masked speech denoising and prediction framework to pre-train speech representations. Some inputs are artificially simulated to be noisy or overlapped with masks, and the model predicts pseudo-labels of the original speech on the masked region.
CCC-Wav2vec 2.0 \cite{ccc-wav2vec} introduces an augmentation of the original sample and uses its representations to add an extra cross-contrastive loss to the Wav2vec 2.0 \cite{w2v} framework. Robust Data2vec \cite{robust_data2vec} improves the noise robustness of the Data2vec \cite{data2vec} by allowing the model to have consistent predictions for both original and noisy speech.
These studies mainly aim to make the SSL model more robust to augmentations, which in turn helps learn better representations.

While augmentation-based approaches are certainly helpful to improve the SSL model's robustness against the data variation, they may not consider randomness in the model architecture.
For example, most recent SSL models are based on the Transformer 
\cite{transformer} structure, which has intrinsic randomness due to multiple dropout variants, such as standard dropout \cite{dropout} for each module, LayerDrop \cite{layerdrop, stochastic_depth}, etc.
During the pre-training stage, these dropout variants randomly discard a portion of layers or neurons, selecting a random \emph{sub-model} at each iteration.
However, when fine-tuning on a downstream task, all or a part of the pre-trained SSL model's parameters are commonly frozen \cite{freeze1,freeze2,freeze3,freeze4,freeze5,lighthubert}, causing inconsistency between the pre-training and fine-tuning.
For ease of understanding, assuming that we freeze all of the SSL model's parameters for the downstream task, the pre-training is performed on a \emph{sub-model} that includes sources of randomness.
In contrast, the fine-tuning is conducted on a single \emph{full model} without randomness. This difference in the amount of randomness between the pre-training and fine-tuning can create a gap between the two stages, leading to performance degradation when applying the SSL model to downstream tasks.

To address this issue, one possible solution is to encourage the sub-models to produce consistent outputs, regardless of the sources of randomness.
If the sub-models are less affected by the randomness, the gap between the sub-model (pre-training) and the full model (fine-tuning) can be alleviated.
Previous studies \cite{rdrop, fd, mean_teacher, mc1} have attempted to improve model-level consistency by regularizing the output predictions of sub-models to be consistent.
However, since these techniques are mainly designed for supervised or semi-supervised settings, it is difficult to directly apply them to fully unsupervised pre-training.
Therefore, it is necessary to design a new consistency regularization framework that can be applied during the pre-training stage of speech SSL.

In this paper, we introduce MCR-Data2vec 2.0, a new pre-training method for improving the model-level consistency of Data2vec 2.0 \cite{data2vec2.0}, which is the current state-of-the-art (SOTA) SSL model. Based on the teacher-student scheme of the Data2vec 2.0 framework, the proposed method randomly samples two different student sub-models, enabling two output variants derived from a single input without additional parameters. 
Due to the dropout variants, the two sub-models are based on different subsets of layers or neurons.
Then, we regularize the outputs of the student sub-models to be consistent with each other and require them to predict the same target representation of the teacher model.
Regularizing the student sub-models to produce similar outputs makes them less affected by randomness.
Thus, MCR-Data2vec 2.0 can effectively reduce the gap between the pre-training and fine-tuning stages, improving the overall quality of the SSL model.

\begin{figure*}[t]
    \centering
        \includegraphics[height=6cm]{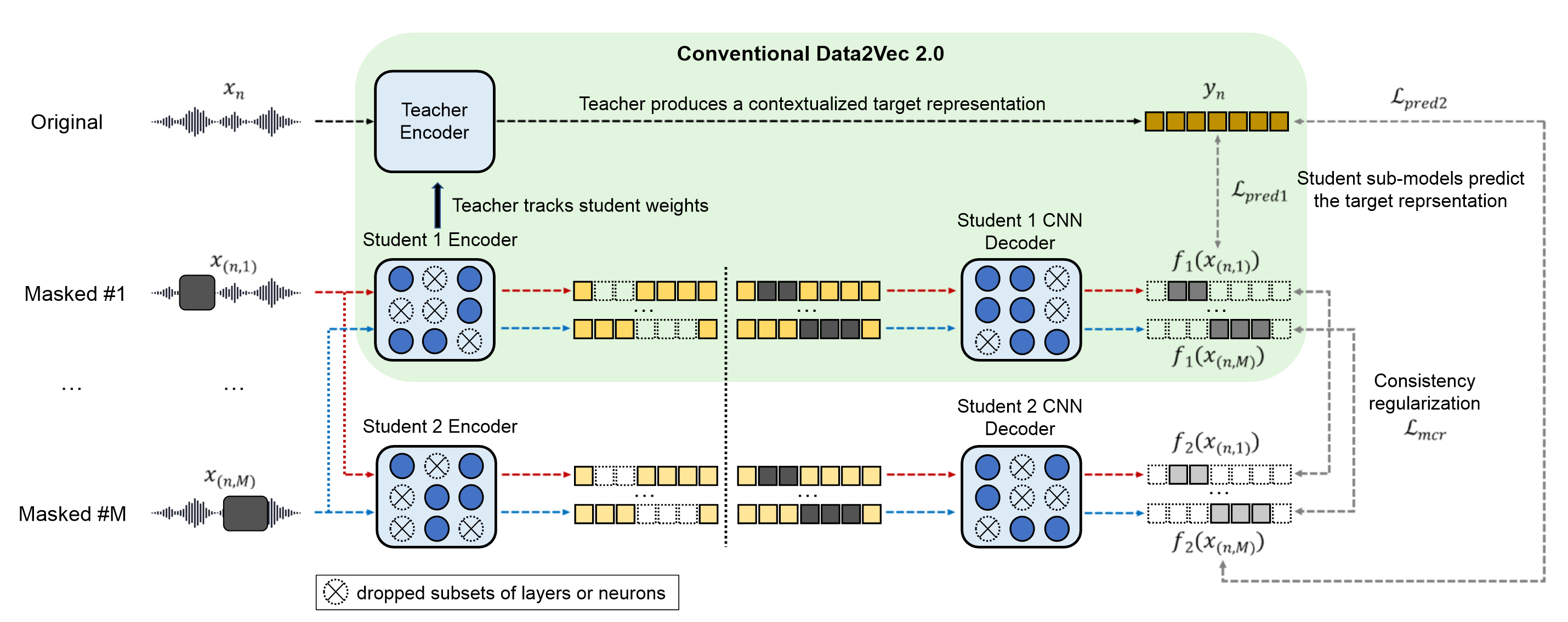}
    \caption{An overview of MCR-Data2vec 2.0. In the student encoder and decoder, a crossed-out dotted circle represents dropped subsets of layers or neurons due to the dropout variants. First, a contextualized representation $y_{n}$ is generated from the teacher model based on the unmasked training sample $x_{n}$.
    The teacher model's weights are a moving average of the student model's weights. 
    MCR-Data2vec 2.0 randomly samples two different student sub-models, resulting in two predictions $f_{1}(x_{(n,m)})$ and $f_{2}(x_{(n,m)})$ for the same masked input $x_{(n,m)}$.
    The masked portions of the training sample are not encoded \cite{data2vec2.0}.
    To improve model-level consistency, the outputs from the student sub-models are regularized to be consistent ($\mathcal{L}_{mcr}$), and the student sub-models predict the same contextualized target representation for various masked versions of the training example ($\mathcal{L}_{pred1}$ and $\mathcal{L}_{pred2}$). $\mathcal{L}_{pred1}$ corresponds to the training objective of the original Data2vec 2.0.
    }
  \label{overview}
\end{figure*}

Through extensive experiments on SUPERB \cite{s3prl, s3prl-sg}, we confirm that the MCR-Data2vec 2.0 effectively improves the SSL model's robustness and generalization ability.
It achieves SOTA performance on multiple subtasks of SUPERB, including Phoneme Recognition (PR), Automatic Speech Recognition (ASR), Keyword Spotting (KS), Intent Classification (IC), Slot Filling (SF), Emotion Recognition (ER).
Furthermore, MCR-Data2vec 2.0 yields considerable performance improvements for speaker and generation-related tasks compared to the original Data2vec 2.0.

To summarize, the main contributions of this paper are:
\begin{itemize}
\item We propose MCR-Data2vec 2.0, a new consistency regularization framework designed to improve the recent Data2vec 2.0.
\emph{To the best of our knowledge, this is the first attempt to apply the model-level consistency regularization during the pre-training of speech SSL.}

\item By regularizing the outputs of student sub-models to be consistent, the proposed method can reduce the gap between the pre-training and fine-tuning stages.

\item Through experimental results on SUPERB, we validate that MCR-Data2vec 2.0 significantly improves the overall quality of Data2vec 2.0. Compared to recent SSL models, our framework achieves promising results on various downstream tasks.
\end{itemize}

\section{Background: Data2vec 2.0}
As aforementioned, we use Data2vec 2.0 \cite{data2vec2.0} as our backbone model since it is one of the most recent SOTA SSL models.
Before we describe the proposed method, it might be beneficial to review some properties of Data2vec 2.0.

\subsection{Teacher-student Setup}
Similar to Data2vec \cite{data2vec}, Data2vec 2.0 follows the teacher-student scheme, where the weights of the teacher $\Delta$ are an exponentially moving average (EMA) of the student encoder $\theta$ \cite{ema}: $\Delta \leftarrow \tau \Delta + (1-\tau) \theta$.
The parameter $\tau$ linearly increases from a starting value $\tau_{0}$ to a
final value $\tau_{e}$ over $\tau_{n}$ updates, after which the value is kept
constant \cite{data2vec, data2vec2.0}.

\subsection{Contextualized Target Prediction}
\label{data2vec_contextualized}
As shown in Figure \ref{overview} (green highlight), Data2Vec 2.0 is based on multi-mask training that considers $M$ different masked versions of the training sample, similar to Masked Autoencoders (MAE) \cite{mae}.
Given a training dataset $\mathcal{D}=\{x_1, x_2, ..., x_N\}$ where $N$ is the number of training data samples, each sample $x_n$ is masked $M$ times. The $m$-th masked version of sample $x_n$ is represented as $x_{(n,m)}$. $y_{n}$ denotes a contextualized target representation from the teacher model based on the original sample $x_{n}$.

In the Data2vec 2.0, the teacher model consumes the unmasked training sample $x_n$ to produce the target representation $y_{n}$.
Specifically, the output of the top $K$ blocks of the teacher model is averaged to construct $y_{n}$.
This target representation is contextualized since the teacher model uses a self-attention mechanism in the Transformer architecture \cite{transformer}.
Each masked version of sample is fed into the student model, which predicts the same target representation $y_{n}$ for the different masked versions.
When passing the masked sample $x_{(n,m)}$, the student does not encode masked tokens to further improve the model efficiency.
During the pre-training, Data2vec 2.0 is trained to minimize the $l_{2}$ loss ($\mathcal{L}_{pred1}$ in Figure \ref{overview}) between the student's prediction and the contextualized target representation.

\section{MCR-Data2vec 2.0}

In this section, we introduce our MCR-Data2vec 2.0 framework and the training algorithm. 
Based on Data2vec 2.0, we propose a new model-level consistency regularization to improve the pre-training stage of SSL. The overall framework of our regularization method is shown in Figure \ref{overview}.

\subsection{Student Sub-model Sampling}
Unlike Data2vec 2.0, MCR-Data2vec 2.0 samples two different student sub-models $f_{1}$ and $f_{2}$ from a full single student model $f$. 
This sub-model sampling allows two output variants to be derived from a single input without requiring additional parameters or model structure changes.

Although it is possible to use multiple sub-models for training, this can be computationally expensive and time-consuming. 
We experimentally confirm that MCR-Data2vec 2.0 can achieve considerable improvements using only two sub-models.

\begin{table*}[t]
\centering
{\fontsize{7.5}{8}\selectfont
\addtolength{\tabcolsep}{-0.4em}
\begin{tabular}{lcccccccccccccc}
\hline \noalign{\vskip 0.02in}
\multirow{3}{*}{Method}   & \multirow{3}{*}{\# Params}             & \multicolumn{4}{c}{Content}                                        & \multicolumn{3}{c}{Semantics}                                      & ParaL                & \multicolumn{2}{c}{Speaker}                 & \multicolumn{3}{c}{Generation}                                     \\ \noalign{\vskip 0.02in} \cline{3-15} \noalign{\vskip 0.02in}
                                &       & PR                   & ASR                  & KS                 &QbE  & IC                   & \multicolumn{2}{c}{SF}                      & ER                   & SID                  & ASV                  & \multicolumn{2}{c}{SE}                      & SS                   \\ \noalign{\vskip 0.02in} \cline{3-15} \noalign{\vskip 0.02in}
                                 &      & PER $\downarrow$                 & WER $\downarrow$                 & Acc $\uparrow$   & MTWV $\uparrow$               & Acc $\uparrow$                   & F1 $\uparrow$                   & CER $\downarrow$                 & Acc $\uparrow$                   & Acc $\uparrow$                   & EER $\downarrow$                 & PESQ $\uparrow$                  & STOI $\uparrow$                  & SI-SDRi $\uparrow$               \\ \noalign{\vskip 0.02in} \hline
\noalign{\vskip 0.02in} 
HuBERT \cite{hubert}             &  94.70 M                   & 5.41                 & 6.42                 & 96.30     & 0.0736           & 98.34                & 88.53                & 25.20                 & 64.92                & 81.42                & {\ul 5.11}                 & 2.58                 & 93.90                 & 9.36                 \\
Wav2vec 2.0  \cite{w2v}    &  95.04 M                     & 5.74                 & 6.43                 & 96.23       & 0.0233         & 92.35                & 88.30                 & 24.77                & 63.43                & 75.18                & 6.02                 & 2.55                 & 93.90                 & 9.77                 \\
WavLM \cite{wavlm}       &      94.70 M                     & 4.84                 & 6.21                 & 96.79   & \textbf{0.0870}             & 98.63                & 89.38                & 22.86                & 65.94                & \textbf{84.51}       & \textbf{4.69}        & 2.58                 & 94.01                & 10.37                \\
Data2vec \cite{data2vec}    &    93.75 M                       & 4.69                 & 4.94                 & 96.56     & 0.0576          & 97.63                & 88.59                & 25.27                & 66.27                & 70.21                & 5.77                 & 2.96                 & 94.83                & 9.78                 \\
CCC-Wav2vec 2.0 \cite{ccc-wav2vec}   &  95.04 M                    & 5.95                 & 6.30                  & 96.72   & 0.0673             & 96.47                & 88.08                & 24.34                & 64.17                & 72.84                & 5.61                 & \textbf{3.06}        & \textbf{94.94}          & \textbf{10.86}       \\
\noalign{\vskip 0.02in} 
\hline
\noalign{\vskip 0.02in}
\multicolumn{1}{l}{\emph{Our implementation}} & \multicolumn{1}{l}{} & \multicolumn{1}{l}{} & \multicolumn{1}{l}{} & \multicolumn{1}{l}{} & \multicolumn{1}{l}{} & \multicolumn{1}{l}{} & \multicolumn{1}{l}{} & \multicolumn{1}{l}{} & \multicolumn{1}{l}{} & \multicolumn{1}{l}{} & \multicolumn{1}{l}{} & \multicolumn{1}{l}{} \\ \noalign{\vskip 0.02in}
Data2vec 2.0 \cite{data2vec2.0}  &  93.78 M                      & {\ul 3.64}           & {\ul 4.81}           & {\ul 96.89}   & {\ul 0.0841}       & {\ul 99.00}          & {\ul 89.67}          & {\ul 22.09}          & {\ul 66.66}          &   81.43                   &      5.59                & 2.98                 & {\ul 94.85}                & 10.41                \\
MCR-Data2vec 2.0 (Ours)  &  93.78 M                  & \textbf{3.37}        & \textbf{4.68}        & \textbf{97.05}   & 0.0595    & \textbf{99.21}       & \textbf{90.04}       & \textbf{21.73}       & \textbf{66.99}       &  {\ul 82.40}                     &  5.36                    & {\ul 3.01}           & \textbf{94.94}          & {\ul 10.62}          \\ \noalign{\vskip 0.02in}  \hline 
\end{tabular}}
\caption{Performance comparison on the SUPERB benchmark. ParaL represents Paralinguistics aspect of speech. All models were trained using 960 hours of LibriSpeech and based on the Base setting, consisting of 12 Transformer blocks. We marked the best in \textbf{bold} and the second-best with {\ul underline}.}
\label{main_result}
\end{table*}

\subsection{Model-level Consistency Regularization}

To train our network, we pass each masked input $x_{(n,m)}$ through the network twice, resulting in two predictions denoted as $f_{1}(x_{(n,m)})$ and $f_{2}(x_{(n,m)})$.
Note that the two forward passes are based on different sub-models due to the dropout variants.
Thus, the predictions $f_{1}(x_{(n,m)})$ and $f_{2}(x_{(n,m)})$ are different for the same masked input $x_{(n,m)}$.

Firstly, the MCR-Data2vec 2.0 requires the student sub-models to predict the same target representation $y_{n}$ of the teacher model.
We minimize the $l_2$ loss between the target representation and the predictions, which can be formulated as:

\begin{align}
\label{pred_loss}
\mathcal{L}_{pred}^{(n,m)} &= \mathcal{L}_{pred1}^{(n,m)} + \mathcal{L}_{pred2}^{(n,m)} \nonumber \\ 
&= (y_{n}-f_{1}(x_{(n,m)}))^{2} + (y_{n}-f_{2}(x_{(n,m)}))^{2} 
\end{align}

where $\mathcal{L}_{pred1}^{(n,m)}$ corresponds to the training objective of the original Data2vec 2.0.

Also, the MCR-Data2vec 2.0 aims to encourage the student sub-models to produce consistent outputs, regardless of randomness in the model architecture.
Since the target of the sub-models is the contextualized representation, both $f_{1}(x_{(n,m)})$ and $f_{2}(x_{(n,m)})$ are continuous.
Thus, we minimize the $l_2$ loss between $f_{1}(x_{(n,m)})$ and $f_{2}(x_{(n,m)})$ to perform the consistency regularization. 
The proposed regularization objective $\mathcal{L}_{mcr}^{(n,m)}$ for the masked sample $x_n^m$ can be computed as follows:

\begin{equation}
\label{loss2}
\mathcal{L}_{mcr}^{(n,m)} = (f_{1}(x_{(n,m)})-f_{2}(x_{(n,m)}))^{2}.
\end{equation}


Although Eq. (\ref{pred_loss}) aims to make two different sub-models predict the same target $y_{n}$, relying solely on this objective function may not sufficiently reduce the variance (difference) between the outputs of the sub-models. 
By minimizing the variance between the outputs of different sub-models via Eq. (\ref{loss2}), we can further improve the model's consistency.

\subsection{Training Objective}
The final objective $\mathcal{L}_{total}^{(n,m)}$ for the sample $x_{(n,m)}$ is given as

\begin{equation}
\mathcal{L}_{total}^{(n,m)} = \mathcal{L}_{pred}^{(n,m)} + \lambda \mathcal{L}_{mcr}^{(n,m)}
\end{equation}

where $\lambda$ is a tunable parameter, and we experimentally set $\lambda$ to 1. 

As mentioned earlier, there is the gap between the pre-training and fine-tuning stages due to the difference in the amount of randomness.
By regularizing the outputs from the two sub-models to be consistent, the sub-models are less affected by the randomness.
Thus, MCR-Data2vec 2.0 can effectively reduce the inconsistency between the pre-training and fine-tuning, improving the overall quality of the SSL model.

\section{Experimental Setup}
\subsection{Pre-training Setup}
We conducted experiments using the Base model configuration, which includes 12 Transformer blocks, 768-dimensional hidden states, and 8 attention heads, resulting in a total of 93.78 M parameters for MCR-Data2vec 2.0.
To pre-train the Data2vec 2.0 Base and MCR-Data2vec 2.0 Base, we used 960 hours of audio from LibriSpeech \cite{librispeech} and trained it for 400K updates on 8 Quadro RTX 8000 48GB GPUs. The models were implemented using fairseq toolkit \cite{fairseq}.
For all other training configurations, we followed the hyperparameters outlined in Data2vec 2.0 \cite{data2vec2.0}.
The multi-masking strategy for the MCR-Data2vec 2.0 was also identical to Data2vec 2.0.

\subsection{Universal Representation Evaluation}
To evaluate the effectiveness of our proposed approach, we conducted experiments using SUPERB \cite{s3prl, s3prl-sg}, a standardized benchmark for pre-trained models in a range of speech tasks. 
We evaluated our model on eleven subtasks of SUPERB, including Phoneme Recognition (PR), Automatic Speech Recognition (ASR), Keyword Spotting (KS), Query by Example Spoken
Term Detection (QbE), Intent Classification (IC), Slot Filling (SF), Emotion Recognition (ER), Speaker Identification (SID), Automatic Speaker Verification (ASV), Speech Enhancement (SE), and Speech Separation (SS).
We followed the guidelines created by SUPERB. Firstly, we employed the same downstream models that were utilized by SUPERB for each task. Secondly, we froze the pre-trained models during the fine-tuning. Finally, the downstream models processed the weighted sum of hidden state results obtained from each layer of the pre-trained model.
For ASR, ASV, SE, and SS, we followed the official configurations of SUPERB. For the other downstream tasks, we followed the fine-tuning hyperparameter settings of WavLM \cite{wavlm}.
During the fine-tuning, we trained the models on a single Quadro RTX 8000 48GB GPU.
\section{Experimental Result}

\subsection{Main Results}
We compared the MCR-Data2vec 2.0 with previous SOTA approaches, including HuBERT \cite{hubert}, Wav2vec 2.0 \cite{w2v}, WavLM \cite{wavlm}, Data2vec \cite{data2vec}, CCC-Wav2vec 2.0 \cite{ccc-wav2vec}, and Data2vec 2.0 \cite{data2vec2.0}.
Table \ref{main_result} summarizes the results on SUPERB benchmark.
Even though Data2vec 2.0 achieved promising results compared to previous SSL methods, MCR-Data2vec 2.0 outperformed the original Data2vec 2.0 in most configurations, except for the QbE task.
For the QbE, the performance of the MCR-Data2vec 2.0 was worse than that of the Data2vec 2.0.
Considering that the HuBERT Large model performed worse than the HuBERT Base model on the QbE task \cite{s3prl}, higher performance on QbE did not always indicate a better SSL model.

As shown in Table \ref{main_result}, the proposed approach yielded SOTA performances for PR, ASR, KS, IC, SF, and ER tasks, while Data2vec 2.0 had the second-best results.
By simply regularizing the output of student sub-models to be consistent, the proposed framework effectively improved the overall quality of the original Data2vec 2.0.
From the results, it is verified that reducing the gap between the pre-training and fine-tuning stages is important in training the SSL model.

Additionally, the proposed method showed significant improvements over the Data2vec 2.0 for speaker and generation tasks, achieving the second-best performances for SID, SE, and SS tasks.
In the case of STOI for SE, MCR-Data2vec 2.0 yielded the best performance along with CCC-Wav2vec 2.0.
It is important to note that WavLM and CCC-Wav2vec 2.0, which performed well on such tasks, benefited from data augmentation for the noisy scenario during the pre-training.
WavLM \cite{wavlm} manually simulated noisy/overlapped speech as inputs, and CCC-Wav2vec 2.0 \cite{ccc-wav2vec} employed three different data augmentations to make the SSL model more robust to augmentations.
However, MCR-Data2vec 2.0 achieved comparable results on speaker and generation tasks without using such data augmentation techniques.
This means that our model-level consistency regularization was supportive in improving the generalization ability and robustness of the SSL model.

\begin{figure}[t]
\vspace{-6mm}
\centering
\subfloat[Data2vec 2.0]{%
  \includegraphics[clip,width=0.8\columnwidth]{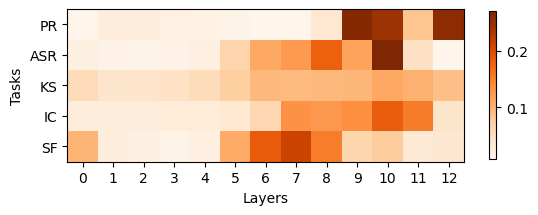}%
}

\vspace{-2mm}
\centering
\subfloat[MCR-Data2vec 2.0]{%
  \includegraphics[clip,width=0.8\columnwidth]{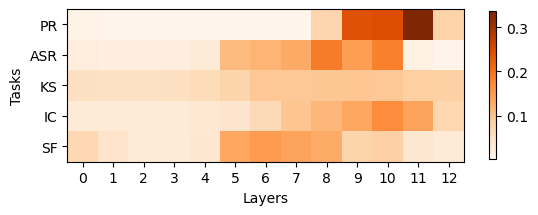}%
}
\vspace{-2mm}
\caption{Weight analysis on content and semantic tasks.}
\vspace{-1mm}
\label{wa}
\end{figure}

\subsection{Weight Analysis}
Following the SUPERB guidelines, we computed a weighted sum of the representations obtained from each layer of the pre-trained model and then fed it to the downstream models.
Since MCR-Data2vec 2.0 showed considerable performance improvements, especially for content and semantic tasks, we analyzed the contribution patterns of the proposed approach. 
Figure \ref{wa} depicts the weights of the different layers of Data2vec 2.0 and MCR-Data2vec 2.0 for PR, ASR, KS, IC, and SF tasks, where the higher weight means that the corresponding layer contributed more to achieving the specific task.
From the results, we found that the patterns of MCR-Data2vec 2.0 were similar to those of Data2vec 2.0.
Even though WavLM \cite{wavlm} reported that the content and semantic information were primarily encoded in the top layers, both SSL models tended to leverage information from both middle and top layers.
In Data2vec 2.0 framework, the use of information from broader layers could be an important factor in achieving better performance for content and semantic tasks.
In addition, we can observe that the contribution patterns of MCR-Data2vec 2.0 were not predominantly concentrated on a specific layer compared to those of Data2vec 2.0.
For example, for the ASR task, Data2vec 2.0 heavily relied on the 10th layer, whereas the patterns of MCR-Data2vec 2.0 were more evenly distributed across layers 5 to 10. Similarly, for the SF task, Data2vec 2.0 predominantly assigned weights to the 6th and 7th layers, while MCR-Data2vec 2.0 utilized information more uniformly across layers 5 to 8.

\subsection{$\mathcal{L}_{pred1}$ Loss Curve}
\begin{figure}[t]
\vspace{-7mm}
    \centering
        \includegraphics[height=3.5cm]{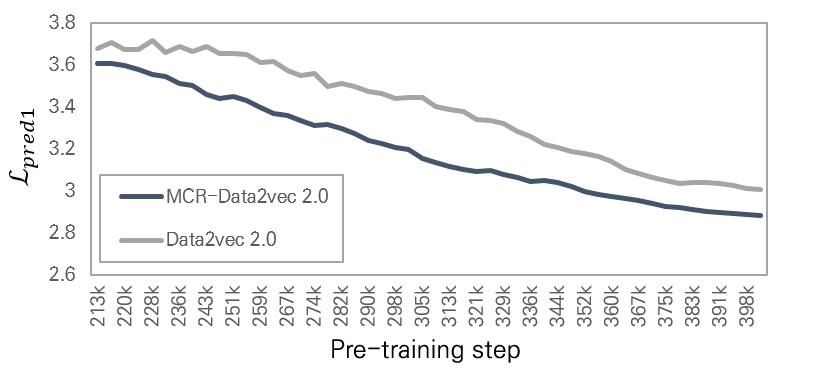}
        \vspace{-3mm}
    \caption{$\mathcal{L}_{pred1}$ loss curves of Data2vec 2.0 and MCR-Data2vec 2.0, where the total number of updates is 400k.
    }
    \vspace{-1mm}
  \label{losscurve}
\end{figure}

To check the effectiveness of the proposed regularization term, we analyzed the pre-training loss function $L_{pred1}$ (in Eq (\ref{pred_loss})).
$L_{pred1}$ measures the distance between the student sub-model's prediction and the target representation of the teacher model.
It is important to note that, as shown in Figure \ref{overview}, both Data2vec 2.0 and MCR-Data2vec 2.0 used $L_{pred1}$ during the pre-training stage.
The loss curves for $\mathcal{L}_{pred1}$ are shown in Figure \ref{losscurve}.
The results showed that $\mathcal{L}_{pred1}$ of MCR-Data2vec 2.0 significantly decreased compared to that of Data2vec 2.0.
This indicates that the proposed regularization improved the student model's ability to predict the target representation of the teacher model.

\section{Conclusion}
We proposed a novel pre-training method, MCR-Data2vec 2.0, to improve the model-level consistency of speech SSL. The proposed framework could reduce the gap between the pre-training and fine-tuning stages while improving the overall quality of the SSL model. From experimental results on the SUPERB benchmark, it is verified that MCR-Data2vec 2.0 achieved promising performance, outperforming recent SSL models.


\newpage
\bibliographystyle{IEEEtran}
\bibliography{mybib}

\end{document}